\documentclass[12pt]{article}
\usepackage{epsfig} 
\newskip\humongous \humongous=0pt plus 1000pt minus 1000pt

\newif\ifdtup

\def\etal{\hbox{\it et al.}} 

\def\VEV#1{\left\langle #1\right\rangle}

\def\pr#1{#1^\prime}

\def\beq{\begin{equation}}
\def\eeq{\end{equation}}

\def\beqn{\begin{eqnarray}}
\def\eeqn{\end{eqnarray}}
\relax

\catcode`\@=11

\@addtoreset{equation}{section}
\def\theequation{\thesection.\arabic{equation}}

\def\@normalsize{\@setsize\normalsize{15pt}\xiipt\@xiipt
\abovedisplayskip 14pt plus3pt minus3pt%
\belowdisplayskip \abovedisplayskip
\abovedisplayshortskip \z@ plus3pt%
\belowdisplayshortskip 7pt plus3.5pt minus0pt}

\def\small{\@setsize\small{13.6pt}\xipt\@xipt
\abovedisplayskip 13pt plus3pt minus3pt%
\belowdisplayskip \abovedisplayskip
\abovedisplayshortskip \z@ plus3pt%
\belowdisplayshortskip 7pt plus3.5pt minus0pt
\def\@listi{\parsep 4.5pt plus 2pt minus 1pt
     \itemsep \parsep
     \topsep 9pt plus 3pt minus 3pt}}

\@twosidetrue

\relax

\catcode`@=12

\catcode`\@=11

\def\section{\@startsection{section}{1}{\z@}{3.5ex plus 1ex minus
   .2ex}{2.3ex plus .2ex}{\large\bf}}

\def\thesection{\arabic{section}}

\def\appendix{\setcounter{section}{0}
 \def\thesection{APPENDIX \Alph{section}:}
 \def\theequation{\Alph{section}.\arabic{equation}}}

\def\ps@headings{\def\@oddfoot{}\def\@evenfoot{}
\def\@oddhead{\hbox{}\hfill
 \makebox[.5\textwidth]{\raggedright\ignorespaces --\thepage{}--
 \hfill {}}}  
\def\@evenhead{\@oddhead}
\def\subsectionmark##1{\markboth{##1}{}}
}

\ps@headings

\catcode`\@=12

\def\figcap{\section*{Figure Captions\markboth
 {FIGURECAPTIONS}{FIGURECAPTIONS}}\list
 {Fig. \arabic{enumi}:\hfill}{\settowidth\labelwidth{Fig. 999:}
 \leftmargin\labelwidth
 \advance\leftmargin\labelsep\usecounter{enumi}}}
 \relax
\def\tablecap{\section*{Table Captions\markboth
 {TABLECAPTIONS}{TABLECAPTIONS}}\list
 {Table \arabic{enumi}:\hfill}{\settowidth\labelwidth{Table 999:}
 \leftmargin\labelwidth
 \advance\leftmargin\labelsep\usecounter{enumi}}}
 \relax
\def\reflist{\section*{References\markboth
 {REFLIST}{REFLIST}}\list
 {[\arabic{enumi}]\hfill}{\settowidth\labelwidth{[999]}
 \leftmargin\labelwidth
 \advance\leftmargin\labelsep\usecounter{enumi}}}
 \relax

\catcode`\@=11

\relax

\relax
\def\pl#1#2#3{{\it Phys. Lett. }{\bf #1}(19#2)#3}
\def\zp#1#2#3{{\it Z. Phys. }{\bf #1}(19#2)#3}

\def\pr#1#2#3{{\it Phys. Rev. }{\bf #1}(19#2)#3}
\def\np#1#2#3{{\it Nucl. Phys. }{\bf #1}(19#2)#3}

\relax

\catcode `@ 11
\def\biblabel#1{\if@filesw\immediate
\write\@auxout{\string\bibcite{#1}{\the\value{\@listctr }}}\fi}
\catcode `@ 12

\newcommand{\ccaption}[2]{
  \begin{center}
    \parbox{0.85\textwidth}{
      \caption[#1]{\small\it {#2}}}
  \end{center}    }
\def    \be             {\begin{equation}}
\def    \ee             {\end{equation}}
\def    \ba             {\begin{eqnarray}}
\def    \ea             {\end{eqnarray}}

\def    \=              {\;=\;}
\def    \frac           #1#2{{#1 \over #2}}

\def \as   {\ifmmode \alpha_s \else $\alpha_s$ \fi}

\def\b0{b_0}

\def \mt   {\ifmmode m_{\rm t} \else $m_{\rm t}$ \fi}

\def \to   {\mbox{$\rightarrow$}}
\newcommand     \MSB            {\ifmmode {\overline{\rm MS}} \else
                                 $\overline{\rm MS}$  \fi}

\newcommand\hepph[1]{{\tt hep-ph/#1}}

\begin{document}
\begin{titlepage}
\nopagebreak
{\flushright{
        \begin{minipage}{4cm}
        CERN-TH/97-92  \hfill \\
        IFUM 566/FT \hfill \\
        hep-ph/9705295\hfill \\
        \end{minipage}        }

}
\vfill
\begin{center}
{\LARGE 
{ \bf \sc  Next-to-Leading-Order Corrections to \\ \vskip 0.2cm
           Momentum Correlations in $Z^0\,\to\, b\bar b$  }}
\vskip .5cm
{\bf Paolo NASON\footnote{On leave of absence from INFN, Milan, Italy},}
\\
\vskip 0.1cm
{CERN, TH Division, Geneva, Switzerland} \\
\vskip .5cm
{\bf Carlo OLEARI}
\\
\vskip .1cm
{Dipartimento di Fisica, Universit\`a di Milano and INFN, Milan, Italy}
\end{center}
\nopagebreak
\vfill
\begin{abstract}
  We recently completed a calculation of the process
  $e^+e^-\,\to\,Q\bar{Q}+X$,
  where $Q$ is a heavy quark, at order ${\cal O}(\as^2)$.
  As a first application of this calculation we compute
  the momentum correlations of $b\bar b$ pairs at next-to-leading-order.
 This quantity is interesting since it may affect the determination
  of $R_b$ as measured in $Z^0$ decays. We find that the next-to-leading
  corrections are of moderate size, thus confirming the conclusions
  that can be drawn from the leading-order calculation.
\end{abstract}
\vskip 1cm
CERN-TH/97-92 \hfill \\
May 1997 \hfill
\vfill
\end{titlepage}
\section{Introduction}
Radiative corrections to jet production in $e^+e^-$ annihilation
have been known for a long time \cite{ERT,VGO,FKSS}. Previous calculations
were, however, performed for massless quarks. In most practical applications
this is sufficient, since at relatively low energy, the $b$ fraction
is strongly suppressed, and at high energy (i.e. on the $Z^0$ peek
and beyond) mass effects are presumably suppressed. Nevertheless
there are several reasons why a next-to-leading-order calculation is desirable.
First of all, at sufficiently high energies, top pairs will be produced,
and mass effects there are very likely to be important. A second important
reason is to understand the relevance of mass corrections due to
bottom production to the determination of $\as$ from event shape variables.
As a third point, quantities such as the heavy flavour momentum correlation
\cite{NO}, although well defined
in the massless limit, cannot be computed using the massless results
of refs.~\cite{ERT,VGO,FKSS}.

\begin{figure}[htb]
\centerline{\epsfig{figure=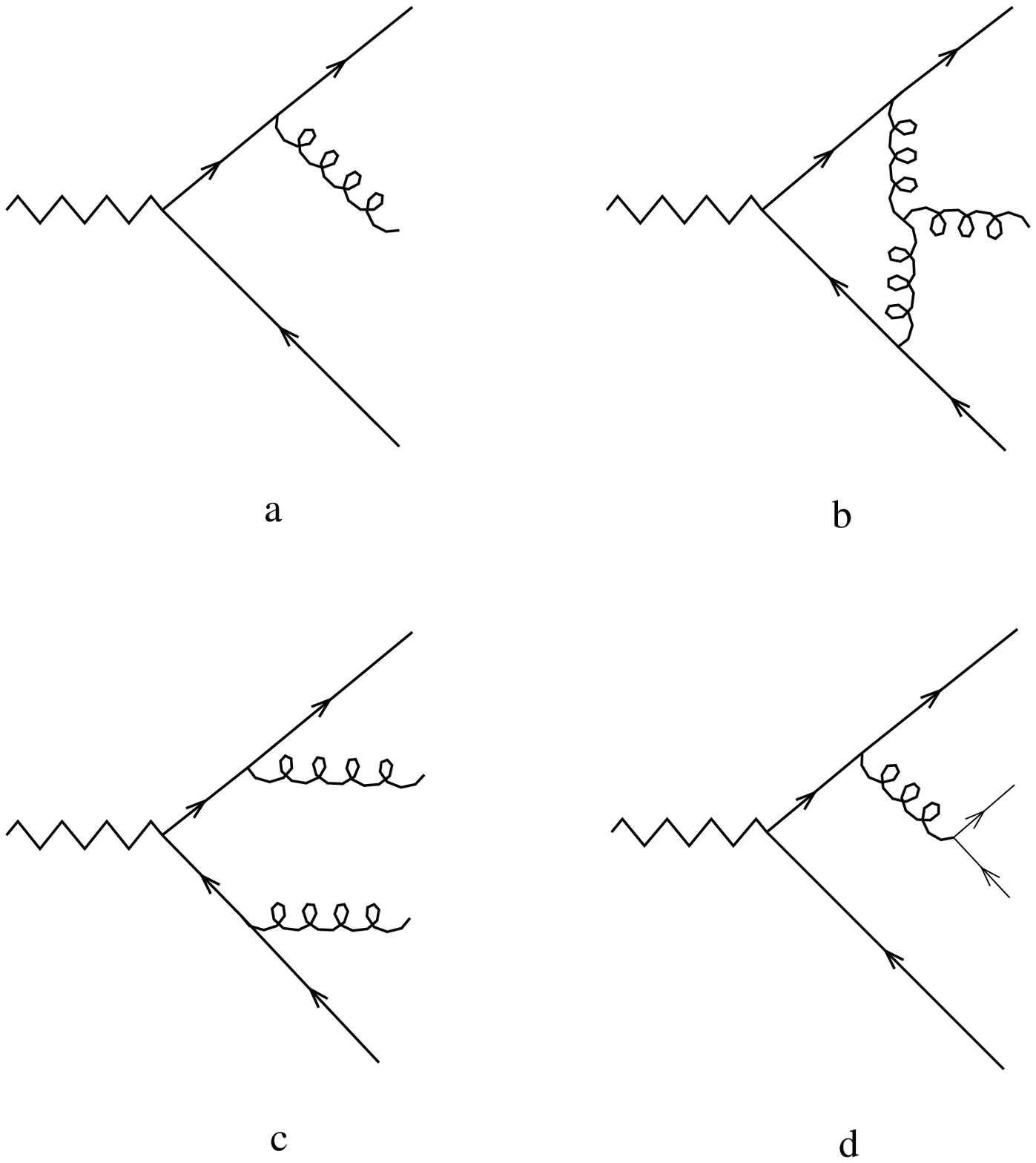,width=0.7\textwidth,clip=}}
\ccaption{}{ \label{fig:graphs}
Some of the diagrams contributing to the process $Z\,\to\, Q\bar Q+X$:
a Born graph (a), a virtual graph (b), a real
emission graph (c), and a real emission graph with light quarks
in the final state (d). }
\end{figure}

We have recently completed a next-to-leading-order calculation of the
heavy-flavour production cross section in $e^+e^-$ collisions,
including quark mass effects.
As an illustration, in fig.~\ref{fig:graphs} we show the Feynman diagrams
for a Born term (a),
a virtual correction term (b), and two real next-to-leading
contributions (c,d).
Next-to-leading corrections arise from the interference
of the virtual graphs with the Born graphs, and from the square
of the real graphs.
Observe that we always deal with the cross section for the production
of the heavy quark pair plus the emission
of at least one extra particle (i.e. a gluon or a quark).
The inclusion of virtual graphs with only a $Q\bar{Q}$ pair in the
final state is not needed if one only computes 3-jet related quantities.
Furthermore, by using available results on the total heavy flavour cross
section at order $\as^2$ (see \cite{KUHN} and references therein), 
one can avoid altogether to compute this kind of virtual graphs.

Virtual graphs, besides the usual ultraviolet divergences, which are removed
by renormalization, have also infrared and collinear divergences.
These cancel when suitable infrared safe final-state
variables are considered. Our treatment of the infrared cancellation
is such that the final result is expressed as a partonic event
generator, in which pairs of weighted correlated events are produced.
Any shape variables can be computed for each generated event.
Only infrared-safe variables give rise to a finite distribution.
No arbitrary cutoff is needed in this calculation in order to implement
the cancellation of virtual and real infrared divergences.
Therefore, one does not have to worry about taking the limit for a vanishing
soft cutoff.
Thus, this method is similar to the one of ref.~\cite{YellowBook},
which was used there to compute a large class of shape variable distributions
for the LEP experiments.

We have verified that our calculation has the correct massless limit,
by comparing our result with that of the massless calculation
of ref.~\cite{YellowBook} for all the shape variables listed there,
and for the moments of the energy--energy correlation, proposed as
a benchmark in ref.~\cite{LEP2}.

Very recently, two calculations have appeared in the literature
that address the same problem \cite{Rodrigo,Bernreuther}.
They both use a slicing method in order to deal with infrared divergences.
We were able to perform a partial comparison of our result with that of
ref.~\cite{Rodrigo}, and found satisfactory agreement.
In the older work of ref.~\cite{Ballestrero},
a calculation of the process $e^+e^-\,\to\, Q\bar{Q}gg$ has been given,
but virtual corrections to the process $e^+e^-\,\to\, Q\bar{Q}g$ were
not included.
In ref.~\cite{Magnea}, the NLO corrections to the production
of a heavy quark pair plus a photon are given, including both real and virtual
contributions. 

We will give full details of our calculation in a future publication.
In the present letter we would like to present a first application,
which could not be dealt with using the massless results.
This is the average
momentum correlation $r$ defined as
\beq\label{defr}
r=\frac{\VEV{x_1 x_2}-\VEV{x}^2}{\VEV{x}^2}\,,
\eeq
where $x_{1(2)}$ are the Feynman $x$ of the produced $B(\bar{B})$ mesons,
$x_{1(2)}=p_{B(\bar{B})}/p^{(\rm max)}_B$, and
$\VEV{x}=\langle x_{1(2)} \rangle$.
In ref.~\cite{NO} $r$ is computed at leading order
in the strong coupling constant. The interest in this quantity lies
in the fact that it may affect the determination of $R_b$ \cite{Rbpapers},
when the tagging efficiency for $B$ mesons is determined by looking at both the
single-tag and the double-tag events. A simplified description
of how this works is given in ref.~\cite{NO}.
It is convenient to rewrite $r$ in the following way
\beq\label{betterr}
r=\frac{\VEV{(1-x_1) (1-x_2)}-\VEV{1-x}^2}{\VEV{x}^2}\,.
\eeq
We notice that, thanks to the above form, the contribution of
virtual graphs with only a $Q\bar{Q}$ pair in the final state
is not needed for this calculation.
Defining now
\beqn
\VEV{(1-x_1) (1-x_2)} &=& \frac{\as}{2\pi}\,b +
       \left(\frac{\as}{2\pi}\right)^2 \!c
+ {\cal O}(\as^3)\,, \nonumber \\
\VEV{1-x} &=& \frac{\as}{2\pi}\, a\,+\,{\cal O}(\as^2)\,,
\eeqn
where $\as$ is the \MSB, 5-flavour coupling constant evaluated
at a scale $\mu$ equal to the annihilation energy $Q$.
We have
\beq
r=\frac{\as}{2\pi}\, b + \left(\frac{\as}{2\pi}\right)^2
    \left(c+2ab-a^2\right)+ {\cal O}(\as^3)\,.
\eeq

In the particular case of the computation of the average momentum correlation,
we have treated as massive the heavy quark directly produced from the
electroweak vertex. Secondary heavy quarks produced by gluon splitting
have been treated as light. Similarly, we have not included graphs
in which a light flavour is produced by the electroweak vertex,
and the heavy quark pair is produced by a gluon. Although the inclusion
of these contributions does not pose any particular problem, we believe
that this quantity is the closest to what is usually defined to be
the correlation induced by hard radiation. Other dynamical correlation
mechanisms, arising for example from the production of four heavy flavours,
or from the indirect (e.g. via gluon splitting) production of a heavy
flavour pair in an event with primary production of light quarks
can be examined separately. Their influence upon the determination
of $\Gamma_b$ depends upon a more detailed specification of
the experimental setup, such as, for example, on whether cuts are applied
that remove soft heavy quarks.

Results for $r$ are displayed in table \ref{tab:logdep}.
\begin{table}[htbp]
  \begin{center}
    \leavevmode
    \begin{tabular}{|c|c|c|c|c|}
      \hline
      $m$    & $a$  & $b$  & $c$ & $r$ \\
      \hline
      1 GeV  &$12.79(1)$&$0.6628(1)$&$155.5(3)$&$0.1055\,\as+0.23(1)\as^2$   \\
      \hline
      5 GeV  &$7.170(2)$&$0.6182(1)$&$50.94(5)$& $0.0984\,\as + 0.213(1)\as^2$ \\
      \hline
      10 GeV &$4.858(1)$&$0.5432(1)$&$26.23(2)$&$0.0865\,\as+0.2004(6)\as^2$ \\
      \hline
    \end{tabular}
    \caption{Mass dependence of the coefficients $a$, $b$, $c$ and $r$.
             The digit in parenthesis (to be taken as zero if not given)
             represents the accuracy of the last digit.}
    \label{tab:logdep}
  \end{center}
\end{table}
From the table it is apparent that
$r$ cannot be computed in the massless limit,
since the quantities $c$ and $a$ are both plagued by collinear divergences.
The coefficients of the expansion of $r$ itself do instead 
converge in this limit.
This fact was discussed at length in ref.~\cite{NO}.
 From the table we also see that the radiative corrections
to $r$ are small.  Assuming $\as(M_Z)=0.118$
(corresponding to the PDG average \cite{PDG}),
we have in leading order $r=0.0984\times\as=0.0116$,
and in next-to-leading order $r=0.0984\times\as+0.213\times\as^2=0.0146$.
Thus, radiative corrections are reasonably under control,
and do not spoil the main
conclusion of the leading-order calculation. With the assumption of
 a rough geometric
growth of the expansion, we can give an estimate of the theoretical
error due to higher orders: $r=0.0146\pm 0.0007$.
We have also computed the quantity $r^\prime$, defined in ref.~\cite{NO}
as
\beq
r^\prime=\frac{\VEV{x_1 x_2\,{\rm cut}(1,2)}-
          \VEV{x}^2}{\VEV{x}^2}
\eeq
where ${\rm cut}(1,2)$ is defined to be 1 if the quark and antiquark
are in opposite hemispheres with respect to the thrust axis,
and zero otherwise. We define
\beq
\VEV{1-x_1-x_2+x_1 x_2\,{\rm cut}(1,2)}=\frac{\as}{2\pi}\,b^\prime +
       \left(\frac{\as}{2\pi}\right)^2\! c^\prime\,+\,{\cal O}(\as^3)\,,
\eeq
and obtain
\beq
r^\prime=\frac{\as}{2\pi}\, b^\prime + \left(\frac{\as}{2\pi}\right)^2
    \left(c^\prime+2ab^\prime-a^2\right)+ {\cal O}(\as^3)\,.
\eeq
The quantities $b^\prime$, $c^\prime$ and $r^\prime$ are given in table
\ref{tab:logdepcut}\footnote{Small numerical discrepancies with ref.\,\cite{NO}
are due to round off errors, and to the inclusion of the electromagnetic
term in the present formulae.}.
\begin{table}[htbp]
  \begin{center}
    \leavevmode
    \begin{tabular}{|c|c|c|c|c|}
      \hline
      $m$    & $b^\prime$  & $c^\prime$ & $r^\prime$ \\
      \hline
      1 GeV  & $0.3800(7)$  & $158.4(3)$ & $0.0605(1)\as + 0.12(1)\as^2$ \\
      \hline
      5 GeV  & $0.3653(6)$  & $50.9(1)$ & $0.0581(1) \as + 0.119(2)\as^2$ \\
      \hline
      10 GeV & $0.3423(5)$  & $25.51(6)$ & $0.0545(1)\as + 0.133(1)\as^2$ \\
      \hline
    \end{tabular}
    \caption{Mass dependence of the coefficients $a$,
             $b^\prime$, $c^\prime$ and $r^\prime$.}
    \label{tab:logdepcut}
  \end{center}
\end{table}
Assuming as before $\as(M_Z)=0.118$ we get $r^\prime=0.0064$ at leading
order, and $r^\prime=0.0083$ at next-to-leading order.

As far as its perturbative expansion in powers of $\as$ is concerned,
the average momentum correlation is a quantity that is well behaved
in perturbation theory, and it is also quite small. Since its effect
is typically of the order of 1\%, one may worry that non-perturbative
effects, of order $\Lambda/Q$ (where $\Lambda$ is a typical hadronic
scale) may compete with the perturbative result.
This is a very delicate problem, since we know very little about
the hadronization mechanism in QCD. In ref.~\cite{NasonWebber},
this problem was addressed in the context of the renormalon approach
to power corrections. It was shown there that (in the renormalon
approach) corrections
to the momentum correlations are at least of order $(\Lambda/Q)^2$,
and thus negligible at LEP energies. Although this result cannot
be considered as a definitive answer to the problem, it is at least
an indication that power corrections to this quantity are small.

\end{document}